# Motion-resolved fat-fraction mapping with whole-heart free-running multiecho gre and pilot tone

Adèle L.C. Mackowiak[1,2,3], Christopher W. Roy[3], Jérôme Yerly[3,4], Mariana B.L. Falcão[3], Mario Bacher[3,5], Peter Speier[5], Davide Piccini[3,6], Matthias Stuber[3,4], Jessica A.M. Bastiaansen[1,2]

[1] Department of Diagnostic, Interventional and Pediatric Radiology (DIPR), Inselspital, Bern University Hospital, University of Bern, Switzerland

[2] Translation Imaging Center (TIC), Swiss Institute for Translational and Entrepreneurial Medicine, Bern, Switzerland

[3] Department of Radiology, Lausanne University Hospital (CHUV) and University of Lausanne (UNIL), Lausanne, Switzerland

[4] Center for Biomedical Imaging (CIBM), Lausanne, Switzerland

[5] Siemens Healthcare GmbH, Erlangen, Germany

[6] Advanced Clinical Imaging Technology (ACIT), Siemens Healthcare AG, Lausanne, Switzerland

**Corresponding author:** Jessica A.M. Bastiaansen, Ph.D

Email: jbastiaansen.mri@gmail.com

Address: Laboratory for Quantitative MR Imaging Sciences (QIS LAB), Department of Diagnostic, Interventional and Pediatric Radiology (DIPR), Inselspital, Bern University Hospital, University of Bern, Switzerland. Freiburgstrasse 3, 3010 Bern, Switzerland

**Word count**

Abstract: 248

Body

Introduction: 555

Methods: 1715

Results: 1400

Discussion & Conclusion: 1256

**Total body: 4926**

# ABSTRACT

## PURPOSE

To develop a free-running 3D radial whole-heart multiecho GRE framework for cardiac- and respiratory-motion-resolved fat fraction (FF) quantification.

## METHODS

Multiecho ($N_{TE}$=8) readouts optimized for water-fat separation and quantification were integrated within a continuous non-ECG-triggered free-breathing 3D radial GRE acquisition. Motion resolution was achieved with Pilot Tone (PT) navigation, and the extracted cardiac and respiratory signals were compared to those obtained with self-gating (SG). After XD-GRASP-based image reconstruction, FF, R2*, and $B_0$ maps, as well as fat and water images were generated with a maximum-likelihood fitting algorithm. The framework was tested in a fat-water phantom and in 10 healthy volunteers at 1.5T using $N_{TE}$=4 and $N_{TE}$=8 echoes. The separated images and maps were compared with a standard free-breathing ECG-triggered acquisition.

## RESULTS

The method was validated *in vivo*, and physiological motion was resolved over all collected echoes. Across volunteers, PT provided respiratory and cardiac signals in agreement (*r*=0.91 and *r*=0.72) with self-gating of the first echo, and a higher correlation to the ECG (0.1% of missed triggers for PT vs 5.9% for SG). The framework enabled pericardial fat imaging and quantification throughout the cardiac cycle, revealing a decrease in FF at end-systole by 11.4±3.1% across volunteers (*P*<0.0001). Motion-resolved end-diastolic 3D FF maps showed good correlation with ECG-triggered measurements (FF bias of -1.06%). A significant difference in free-running FF measured with $N_{TE}$=4 and $N_{TE}$=8 was found (*P*<0.0001 in sub-cutaneous fat and *P*<0.01 in pericardial fat).

## CONCLUSION

Free-running fat fraction mapping was validated at 1.5T, enabling ME-GRE-based fat quantification with $N_{TE}$=8 echoes in 6:15min.

## 1. INTRODUCTION

MRI-derived proton-density fat fraction (PDFF) is considered a robust and reproducible noninvasive measure of fat concentration within the MR research community[1]. Cardiac fat quantification canaid in diagnosing pathologies where adipose cells abnormally develop, within fat depots in dilated cardiomyopathy[2] or within scar tissue of the infracted myocardium leading to an increased risk of arrhythmogenic right ventricular cardiopathy[3,4] and sudden cardiac death[5,6]. Fat fraction (FF) quantification also carries potential to characterize the complex metabolic role of adipose tissues in obesity[7,8] and diabetes[9,10], in which increased amounts of epicardial, pericardial and peri-coronary fat alter the cardiovascular disease risk profile[11,12]. Nevertheless, cardiac fat quantification with MRI is seldom performed in clinical settings, where invasive biopsies remain the standard measurement.

PDFF can be quantified using multi-echo GRE (ME-GRE) MRI sequences that acquire images at different echo times[13]. Multiple echoes are needed to reliably separate the signals in the presence of $B_0$ field inhomogeneities, which confounds fat detection[14]. Dedicated algorithms mitigate the effects of $B_0$[15–18], $T_1$[19,20], $T_2^*$[21], or noise[22], and assume a fixed fat spectral model[23–25]. Accurate fat quantification requires a sufficient number of echoes $N_{TE}$ to resolve the multiple resonance peaks of triglycerides, which lengthens the acquisition time and may limit clinical translation. Because of motion, cardiac PDFF quantifications are typically performed during breath-holds and use triggering devices such as electrocardiograms (ECG). Therefore, such measurements are limited in the number of echoes collected, or restricted in terms of organ coverage. While free-breathing techniques have enabled whole-heart water-fat separation at 1.5T[26,27], 3T[28,29] and 7T[30], they still relied on triggered acquisitions with NTE≤4, and did not focus on quantification. Alternatively, approaches that combine fingerprinting[31–33] or deep learning[34] with ME-GRE have shown promising results for fat quantification with various $N_{TE}$, but still required breath-holding and ECG-triggering with restricted organ coverage. In order to improve spectral resolution with an increased number of echoes while maintaining scan efficiency, alternative motion management techniques are needed.

Recent free-running concepts using uninterrupted 3D acquisitions enable whole-heart free-breathing MRI where ECG time stamps help resolve cardiac motion retrospectively[35]. Free-running sequences have a fixed scan time, improve ease-of-use, and applications range from anatomical imaging[36–38], to coronary angiography[39,40], T1 and T2 mapping[41–44], and flow measurements[45]. Advances in respiratory motion compensation extended with a compressed sensing (CS) reconstruction enabled cardiac- and respiratory-motion-resolved 5D imaging[46,47]. The addition of Pilot Tone (PT) technology[48,49] as an alternative to self-gating (SG) for extracting physiological signals[50] allows sequence-independent motion monitoring. This development could be particularly suitable for long or repeated echo readouts, such as ME-GRE scans, which typically have lower SNR.

To address the challenge of limited organ coverage and restrictions on the number of echoes that can be acquired, a 3D ME-GRE free-running approach was developed for FF quantification. The aim of the study was to combine the strength of motion-resolved cardiac MRI and advanced fat-water decomposition techniques to perform multi-peak fitting of 3D whole-heart ME-GRE data, which are resolved for cardiac and respiratory motion, without needing triggering nor breath-holding. The approach combines 1) an extension of the free-running acquisition to multiecho sampling, 2) PT technology, 3) robust CS reconstruction and 4) a multi-peak fitting routine for fat-water separation and quantification, with 8 echoes. Phantom and healthy volunteer experiments were performed to test whether the proposed free-running FF mapping framework can provide 3D motion-resolved parametric maps of cardiac adipose tissue.

## 2. METHODS

Experiments were performed at 1.5T (MAGNETOM Sola, Siemens Healthcare, Erlangen, Germany) using a 12-channel body array equipped with an integrated PT generator. Volunteers, N=10 (F=5 age [21;31] y.o., BMI [19.1;24.7]) provided their informed consent and the study was approved by the local Ethics Committee.

*2.1 DATA ACQUISITION, RECONSTRUCTION AND POST-PROCESSING FRAMEWORK*

### 2.1.1 Sequence design

A prototype whole-heart 3D radial free-running GRE sequence was modified to incorporate multiple echo readouts for each radial segment using monopolar readout gradients (**Figure 1**). The phyllotaxis k-space trajectory consists of radial segments grouped into interleaved spirals, which are successively rotated by the golden angle[51]. The first radial segment of each interleaf is oriented along the superior-inferior (SI) direction. The multi-echo readout scheme prolongs the time between two subsequent SI segments of the same echo time $TE_i$ (i=1,…,$N_{TE}$) within the echo train (see **Supporting Information Table S1**). The SI segments are used to perfom self-gating (see Section 2.1.3). The ECG signal was recorded during the scan.

An ECG-triggered version of above sequence was acquired, using the same trajectory, with data collection performed during the diastolic resting phase.

### 2.1.2 Acquisition parameters

*In vitro*

A fat-water phantom with 14 vials containing different fat concentrations (see **Supporting Information Figure S1** for details on phantom design and a comparison with a Cartesian acquisition) was scanned using $N_{TE}$=8 and $N_{TE}$=5, to mimic respectively the free-running and ECG-triggered sequences, and with two RF excitation angles α=12° and α=6°. Sequence parameters included: a pixel bandwidth of 890Hz/px, a field-of-view (FOV) size of 290mm$^3$, an interecho spacing of ΔTE=2.05ms, and TE$_1$=1.25ms. The repetition time was TR=17.02ms in free-running and TR=10.87ms with ECG-triggering.

*In vivo*

The prototype free-running sequence was designed with a fixed acquisition time (TA) of 6:15min and isotropic resolution of (2.0mm)$^3$ (**Table 1**). NTE=8 echoes with an interecho spacing ΔTE=2.05ms were collected with a trajectory of 22 segments per interleave. The ECG-triggered protocol had a matching FOV, spatial resolution, RF excitation angle, receiver bandwidth, and a similar TA. The ECG-triggered sequence is heart rate-dependent with an average TA=6:23min. The time available for echo collection is limited by the cardiac resting phase period. Therefore, the trajectory was set to 12 segments and 437 interleaves in the triggered protocol. This enabled the collection of N$_{TE}$=5 echoes while producing a similar phyllotaxis pattern as in the free-running protocol (**Supporting Information Figure S2**), and a slightly higher Nyquist sampling factor (**Table 1**).

| MR acquisition | Free-Running | ECG-triggered |
|---|---|---|
| FOV size [mm$^3$] | 220 x 220 x 220 | 220 x 220 x 200 |
| Spatial resoltion [mm$^3$] | 2.0 x 2.0 x 2.0 | 2.0 x 2.0 x 2.0 |
| RF excitation angle [°] | 12 | 12 |
| Receiver bandwidth [Hz/px] | 893 | 893 |
| TR [ms] | 17.02 | 11.04 |
| N$_{TE}$ | 8 | 5 |
| TE$_1$ / ΔTE | 1.25 / 2.05 | 1.25 / 2.05 |
| TA [min:s] | 6:15 | [5:58; 6:64] |
| N$_{segments}$ x N$_{interleaves}$ | 22 x 1000 | 12 x 437 |
| Nyquist sampling factor [%] | 5.6 | 6.4 |

Table 1 – MR sequence parameters for the free-running and ECG-triggered ME-GRE acquisitions

Sequence parameters were chosen to maximize data collection within a scan time of 6 minutes, with matching echo times (TE) and inter-echo spacing ΔTE for the free-running and ECG-triggered acquisitions.

The Nyquist sampling factor corresponds to the ratio between the k-space lines within one motion bin and the total amount of lines required for a fully-sampled reconstruction, expressed in %.

The proposed free-running sequence induced Specific Absorption Rate (SAR) values ranging from 0.01719 to 0.02197 W/kg in the N=10 volunteers.

### 2.1.3 Physiological signals extraction

The coil-integrated PT generator emits a continuous-wave RF signal that is modulated by physiological motion. PT functions at a frequency outside the MR band (62 MHz), therefore not disturbing image acquisition. The PT data was used to compute one-dimensional respiratory and cardiac signals using principal component analysis (**Figure 1**), which were then used to divide the raw data into 2 respiratory and 10-11 cardiac motion states. Respiratory motion was adressed by selecting the radial views in the bin corresponding to end-expiration (around 40%, as was done previously[52]). Compared to previously published single-echo free-running reconstructions [39,47,50], the proposed multiecho framework uses a lower temporal resolution (90ms per cardiac bin) in order to guarantee an acceptable undersampling factor (**Table 1**) for CS reconstruction, as well as sufficient SNR towards the end of the long echo train.

### 2.1.4 Motion-resolved image reconstruction

The k-space data were sorted into 6D (x-y-z-cardiac-respiratory-echo) matrices according to the motion states and echo number. The first 220 radial segments were discarded from the reocnstruction to eliminate potential transient magnetization effects. The highly undersampled motion-resolved datasets were reconstructed using XD-GRASP[46,53] and the alternating method of multipliers (ADMM)[54]. Regularization using total variation[53] was applied spatially, as well as along two motion-resolved dimensions. No regularization was applied along the echo dimension, so the contrast variations between fat and water remained unaltered. The same regularization, chosen to maintain a good trade-off between visual image quality and motion compression artifacts, were used for all 6D reconstructions (**Supporting Information Figure S3**).

All 6D reconstructions were performed using MatLab R2018b (The Mathworks, Inc., Natick, Massachusetts, USA) on a workstation equipped with 2 Intel Xeon CPUs (Intel, Santa Clara, California, USA), 512GB of RAM, and a NVIDIA Tesla GPU (Nvidia, Santa Clara, California, USA). Reconstruction time was recorded using a built-in MatLab clock.

Respiratory-motion-resolved image reconstruction of the ECG-triggered acquisitions was performed with the same algorithm, with matching regularization where necessary.

### 2.1.5 Fat-water separation and quantitative parameters (FF, $B_0$, R2*) estimation

The 6D datasets were post-processed to compute fat- and water-only 5D (x-y-z-cardiac-respiratory) images as well as parametric maps (FF, R2*, $B_0$) with an iterative graph cut algorithm for fat-water separation[18,55], part of the ISMRM 2012 Fat/Water Toolbox[56]. The algorithm estimates the $B_0$ map and computes water and fat fraction maps through fitting a 6-peak spectral fat model with a single T2* decay component[23,56]. Parameters included: a range of [0;100]Hz for R2* estimation, a range of [-400;400]Hz for the $B_0$ map, a number n=40 of graph cut iterations, and a regularization λ=0.05. A spatial subsampling with factor R=2 was used to accelerate the $B_0$ map estimation.

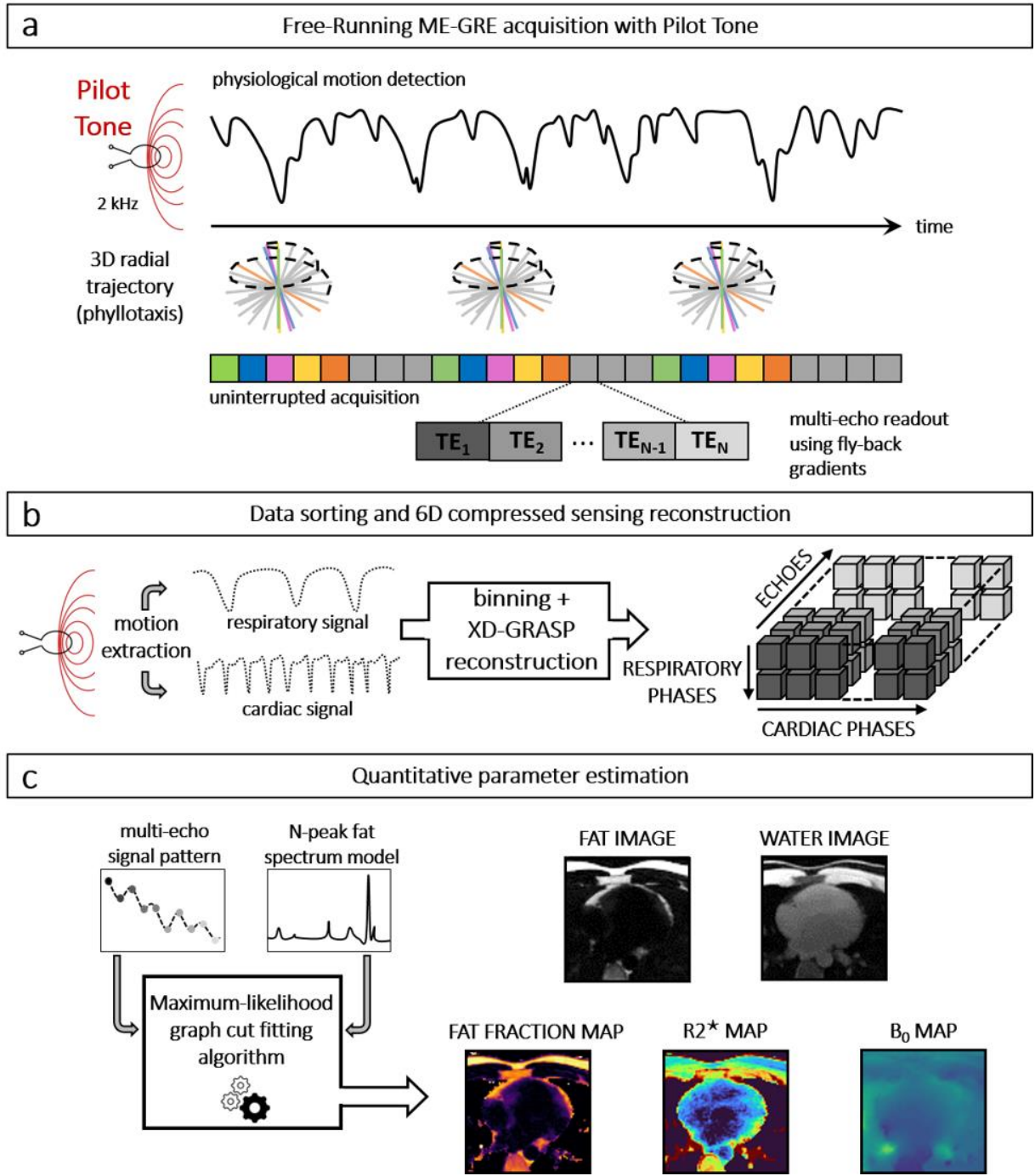

**Figure 1 – Acquisition, reconstruction and post-processing framework**

**a**. The free-running multiecho GRE acquisition uses a 3D radial phyllotaxis trajectory, where each k-space line is repeated $N_{TE}$ times with fly-back gradients. During the whole duration of the uninterrupted data acquisition, the coil-integrated Pilot Tone navigator registers signals at a 2kHz sampling rate, which are modulated by physiological motion.

**b**. Acquired signals are used to bin the MR data into cardiac and respiratory motion states. The XD-GRASP algorithm reconstructs the highly undersampled k-space data into 6-dimensional imaging volume.

**c**. The signal pattern formed along the echo dimension is fed to a fat-water decomposition algorithm, along with a reference fat spectrum, to produce separated fat and water images, as well as fat fraction (FF), water fraction (WF) and main field deviation ($B_0$) maps.

## *2.2 ANALYSIS*

### 2.2.1 Comparison of Pilot Tone and Self-Gating for ME-GRE

Physiological signal extraction using PT was compared to a self-gating (SG) approach. SG uses principal component analysis[57] on the repeated SI projections that encode motion from all active receiver channels. Because 8 subsequent SI projections are acquired, the effect of the choice of SI projection for self-gating of multiecho acquisitions was tested in all volunteers by performing the SG signal extraction from the eight sets of SI projections (labelled SG $TE_i$, i=1,…,8) and comparing it to signals extracted from PT and the reference ECG trace. A Pearson correlation analysis was performed to determine which SG $TE_i$ source provides the closest match to PT and to determine the variability between the different sources. The influence on motion binning was determined by computing a percentage of binning difference to PT. This metric corresponds to the ratio of k-space segments placed into a different bin than the one selected using the PT signal, over the total number of segments. Additionally, a visual comparison was performed on the reconstructed images.

Because SG signals are derived from 3D radial imaging data, they may contain trajectory-made frequency components that are non-physiological and require filtering[47] (**Supporting Information Figure S4**). To determine the amount of trajectory-dependent information embedded within the SG cardiac signals extracted from each TE, a metric of spectral power removal was used. Spectral power removal corresponds to the percentage of spectral density power removed from the original frequency spectrum after the trajectory-dependent frequency component filtering[47]. This metric informs on the impact of gradient delays and eddy currents on the recorded signal.

Cardiac signals extracted from all sources were also compared to the ECG trace. The mean durations of the cardiac cycle were compared to those measured by the ECG. The number of trigger points throughout the free-running acquisition was reported for the ECG and the other sources, and the percentage of missed triggers (w.r.t. ECG triggers) was computed. Moreover, the trigger jitter, i.e. the standard deviation of the difference between the time stamps of the ECG triggers and that of each source, was measured. This last metric informs about the accuracy of cardiac cycle length estimation, as the trigger points of each source are not matched with the ECG trigger time: PT triggering is performed on the local minima of the extracted cardiac signal, while SG triggering is performed on the zero-crossing point (**Figure 4b**). Therefore, the trigger jitter measures a deviation across pairs of associated trigger points between the ECG and each evaluated source.

### 2.2.2 Parametric mapping analysis

To analyze FF maps, regions of interest (ROIs) were drawn in pericardial fat, and sub-cutaneous fat for a static reference. ROIs were drawn based on the visual assessment of the fat-only images. For each tissue type, two ROIs were drawn, both in free-running and in ECG-triggered data. The average FF and standard deviation was computed across cardiac motion states, at expiration. The FF from triggered datasets were compared to those from the diastolic resting phase in the free-running data by linear regression and Bland-Altman analyses.

### 2.2.3 Impact of echo train length

To test the impact of $N_{TE}$ on fat quantification, free-running data were undersampled by selecting only the first four echoes (from $TE_1$ = 1.25 ms to $TE_4$=7.40 ms). The FF maps obtained with $N_{TE}$=4 and $N_{TE}$=8 were compared. The average FF over 10 end-diastolic, expiratory slices in two ROIs in each tissue type were computed. Paired parametric t-tests (GraphPad Prism, San Diego, California, USA) were performed in each tissue to determine statistically significant differences.

### 2.2.4 Impact of T1 bias

Due to the shorter T1 relaxation time of fat compared to water-based tissues, the fat signal measured with GRE imaging is amplified by a factor κ which depends on the RF excitation angle[19,22]. When defining the true fat fraction as $FF_{true}$=F/(W+F) where F and W are the respective amplitudes of fat and water signals, the measured fat fraction $FF_{measured}$ deviates from $FF_{true}$ so that $FF_{measured}$=κF/(W+κF). To evaluate the impact of T1 bias, numerical simulations of the Bloch equations (MatLab 2021a, MathWorks, Natick, Massachusetts, USA) and phantom experiments were performed. The signal evolution of myocardial and fat tissues was simulated for both the free-running (TR=17.02ms) and the ECG-triggered (TR=11.04ms) sequences, for a range of RF excitation angles α∈[1 ;25]°. The relaxation times were T1=996ms and T2=47ms for myocardium[58], and T1=343ms and T2=58ms for sub-cutaneous fat at 1.5T[59]. For the ECG-triggered sequence, a cardiac cycle length of 900ms was assumed. The measured fraction $FF_{measured}$ was plotted as a function of $FF_{true}$ for the choice of RF excitation angle α=12° used in volunteers.

*In vitro*, the same fat-water separation post-processing was performed as described in volunteer experiments. The FF average and standard deviation was measured in phantom vials across 15 slices and compared to reference values obtained with MR spectroscopy (MRS) performed at 9.4T. Following the methodology of Yang *et al*.[20], a T1 bias correction was performed using T1 estimates (T1$_{fat}$=340ms and T1$_{water}$=1350ms measured with inversion recovery turbo-spin-echo MRI[60]) on the separated images, and corrected FF values were also calculated and compared with MRS.

In volunteers, T1 bias correction was performed to adequately compare the free-running and ECG-triggered FF maps: for each sequence, the bias estimated from the simulation experiments described above was used.

### 2.2.5 Blinded scoring

A blinded scoring of the water-only and fat-only images and the FF maps w.r.t. motion resolution, presence of residual sampling artifacts and tissue delineation was performed by two engineers experienced in radial compressed sensing, following a 5-point Likert scale. More information and results can be found in **Supporting Information Table S2**.

## 3. RESULTS

### *3.1 NUMERICAL SIMULATIONS & PHANTOM EXPERIMENTS*

#### 3.1.1 Numerical simulations

A fat signal amplification factor of κ=1.5854 was found for an RF excitation angle of 12° (**Figure 2ab**, dashed curves). For the ECG-triggered sequence, this factor was κ=1.1610 (**Figure 2ab**, plain curves). This translated into different correction curves for the free-running and ECG-triggered sequences, where the largest deviation was measured at 11.4% for FF$_{true}$=44.3% with the free-running sequence, and FF$_{true}$=48.1% with a deviation of 3.8% (**Figure 2b**).

#### 3.1.2 Phantom experiments

Both the free-running and ECG-triggered protocols demonstrated large deviations due to T1 bias. Even with a reduced flip angle approach (α=6°) the bias was not completely eliminated (**Figure 2cd**). After bias correction with known T1 estimates, the regression analysis w.r.t. the MRS controlled FF produced a line described by y=1.02x-0.67 (r$^2$=0.995) for the free-running protocol and y=0.99x+1.73 (r$^2$=0.998) for the ECG-triggered protocol.

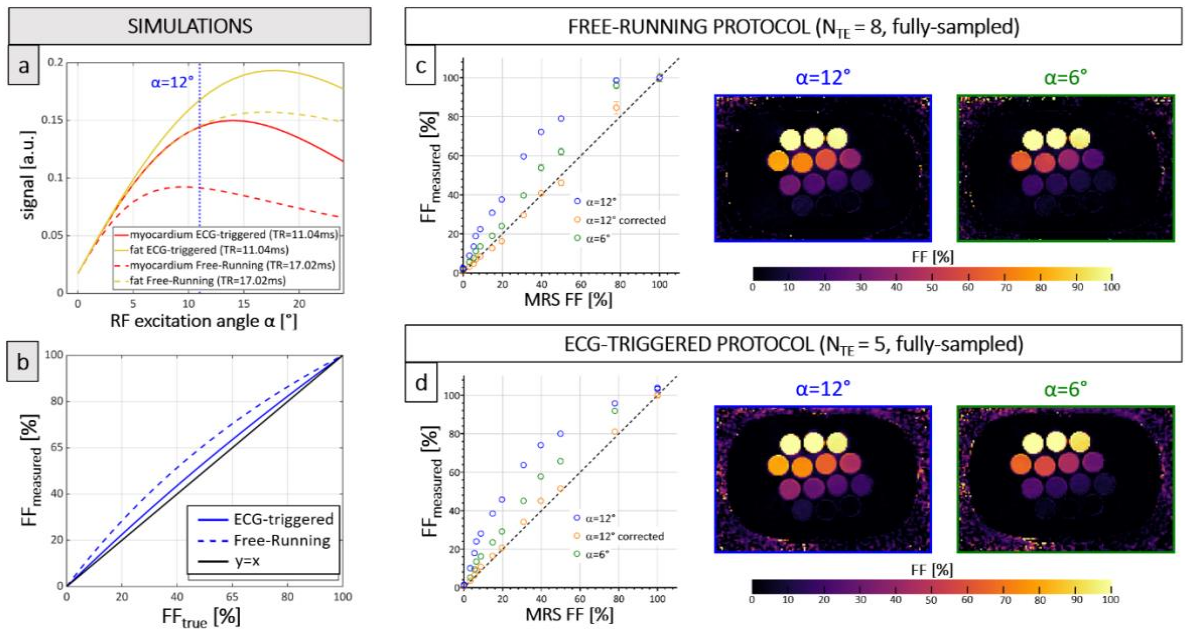

**Figure 2 – Impact of T1 bias in numerical simulations and in a phantom with controlled fat fractions**

Using Bloch simulations, simulated fat and myocardium signals are plotted for different RF excitation angles in a free-running and in an ECG-triggered sequence (**a**). The blue dotted line indicates the expected signal amplitudes in the current study when using an RF excitation angle α=12°. In panel (**b**), the measured FF obtained with α=12° is plotted as a function of the true FF, for the free-running and ECG-triggered sequences.

In phantom experiments, the average FF measured in maps obtained with free-running (**c**) and ECG-triggered (**d**) sequences, with RF excitation angles α=12° (blue), α=6° (green), is shown as a function of ground truth FF measured with MRS. T1bias-corrected values from the α=12° protocols are shown in orange on the graphs (**c**, **d**). Although the reduced flip angle approach reduced the bias of T1 to a certain degree, the correction based on T1 estimates completely eliminated it.

### 3.2 MOTION-RESOLVED MULTI-ECHO IMAGE RECONSTRUCTION

PT navigation successfully extracted both cardiac and respiratory motion in all 10 healthy volunteers, allowing the echo-specific visualization of motion frames corresponding to the expiration phase of the respiratory cycle, and the end-diastolic phase of the cardiac cycle (**Figure 3, Supporting Information S3**). The 10 (11 bins for volunteer V6 who had the longest mean cardiac cycle) cardiac bins had a bin length between [74.5;102.5] ms, with an average bin length of 90.7 ms across volunteers.

The average XD-GRASP reconstruction time for the 10 6D free-running datasets was 5h05min±1h10min. For the 10 5D ECG-triggered datasets, the average reconstruction time on the same workstation was 26min±4min.

Free-running images were in visual agreement with reference ECG-triggered images obtained at the diastolic resting phase in terms of anatomy, and showed for each TE anatomical features that cannot be recovered from heavily undersampled data without CS (**Figure 3**). In addition to signal loss along the echo dimension, additional blurriness at organ interfaces (heart-liver, lung-liver) was also observed at $TE_8$, compared to $TE_1$ and $TE_4$. Despite more data used for each 3D volume reconstructed from the triggered acquisition (6.4% Nyquist against 5.6% for free-running) and the use of the same PT-based respiratory motion correction, signal losses outside the heart were observed in triggered images that were much attenuated in free-running images (**Figure 3**).

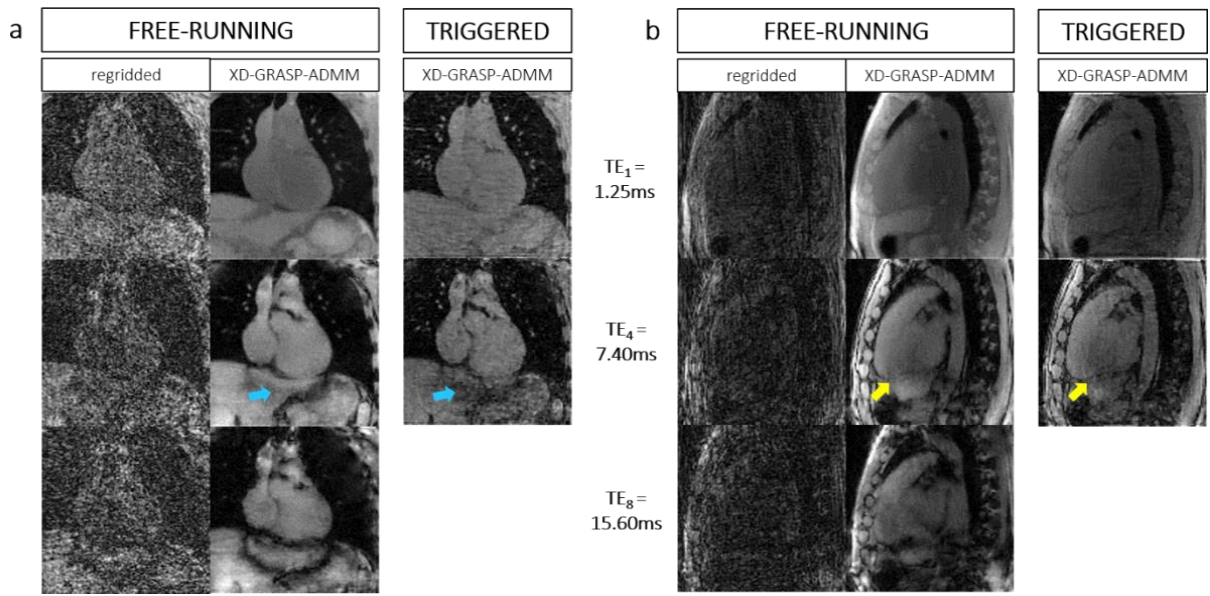

Figure 3 – Multi-echo compressed sensing reconstruction with XD-GRASP-ADMM

The effect of XD-GRASP on the different contrast images (at echo times $TE_1$, $TE_4$ and $TE_8$) collected with the PT free-running ME-GRE acquisition is shown in (**a**) the coronal plane in healthy volunteer V6 and (**b**) the sagittal plane on healthy volunteer V7. The framework allows recovering of the anatomy absent from the highly undersampled regridded data (5.6% Nyquist), whilst preserving the contrast change needed for chemical species separation. In comparison, the ECG-triggered 5-echo acquisition (6.4% Nyquist) reconstructed with the same framework displays a slightly higher visual sharpness (**b**, yellow arrows), but is noisier than the free-running images. The ECG-triggered image at $TE_4$ shows some signal loss and blurriness in the liver that is less visible in the PT free-running data (**a**, blue arrows).

## 3.2 COMPARISON OF PILOT-TONE AND SELF-GATING FOR ME-GRE

Respiratory signals extracted from the SG sources showed good consistency across echoes, with an average Pearson correlation to PT $r≥0.93$ across volunteers (**Table 2**). As illustrated in volunteer V1, the SG respiratory signals from the first two and the last two recorded echoes showed good agreement with PT, despite a smaller peak-to-peak amplitude reported for PT over the respiratory cycles displayed (**Figure 4a**). Higher correlation was found across the volunteers for the first two echoes (Table 2). Only the selection of $TE_8$ as a source tends to increase the deviation from PT binning (18.6% compared to 11% for other sources).

For cardiac motion, variability was observed in the detected cardiac frequency $f_{CARD}$ when a different TE source is used, which is visualised in volunteer V1 over a dozen cardiac cycles in **Figure 4b**. Overall, the correlation to PT was poorer for cardiac than respiratory signals, with the best correlation at *r*=0.72 achieved with SG $TE_1$, and the weakest correlation at *r*=0.36 with SG $TE_7$ across volunteers (**Table 2**). It is worth noting that large standard deviations were observed across volunteers for this metric.

The percentage of spectral power removal decreased with increasing source echo time (**Table 2**). With SG $TE_1$ as source, an average of 92% of frequency components were identified as trajectory-dependent and thus filtered out. However, using the next echo $TE_2$ as source, this number dropped to 60%. A steady reduction was seen with the use of successive TEs, with the lowest number reported for $TE_7$. However, no particular trend across echoes was observed in the percentage of binning difference with respect to PT-based binning (**Table 2**).

The trigger points from the PT cardiac signals had the best match to ECG triggers, with a trigger jitter of (23±14)ms across volunteers and 0.1% of missed triggers. SG $TE_7$ had the largest deviation from both the PT source and the ECG signal across volunteers, with an average trigger jitter as large as (118±102)ms.

| source | % spectral power removed | Pearson's *r* correlation to PT – respiratory | Pearson's *r* correlation to PT – cardiac | % binning difference to PT – respiratory | % binning difference to PT – cardiac | % missed ECG triggers | ECG trigger jitter [ms] |
|---|---|---|---|---|---|---|---|
| SG $TE_1$ | 92 ± 5 | 0.96 ± 0.01 | 0.72 ± 0.32 | 12 ± 2 | 15 ± 4 | 5.9 ± 4.4 | 69 ± 22 |
| SG $TE_2$ | 60 ± 13 | 0.96 ± 0.03 | 0.51 ± 0.24 | 12 ± 2 | 14 ± 4 | 6.8 ± 9.6 | 87 ± 24 |
| SG $TE_3$ | 58 ± 10 | 0.94 ± 0.06 | 0.54 ± 0.31 | 12 ± 2 | 13 ± 4 | 9.0 ± 12.8 | 132 ± 92 |
| SG $TE_4$ | 32 ± 11 | 0.93 ± 0.10 | 0.66 ± 0.26 | 11 ± 2 | 12 ± 4 | 5.4 ± 8.5 | 68 ± 22 |
| SG $TE_5$ | 30 ± 7 | 0.92 ± 0.11 | 0.52 ± 0.32 | 12 ± 2 | 13 ± 4 | 10.8 ± 14.1 | 65 ± 26 |
| SG $TE_6$ | 29 ± 7 | 0.92 ± 0.11 | 0.60 ± 0.29 | 11 ± 1 | 13 ± 4 | 6.9 ± 9.5 | 57 ± 16 |
| SG $TE_7$ | 21 ± 9 | 0.92 ± 0.11 | 0.36 ± 0.31 | 11 ± 2 | 13 ± 4 | 13.9 ± 14.5 | 118 ± 102 |
| SG $TE_8$ | 28 ± 6 | 0.91 ± 0.12 | 0.56 ± 0.30 | 19 ± 3 | 13 ± 3 | 8.5 ± 14.0 | 85 ± 55 |
| Pilot Tone | 0 | 1 | 1 | 0 | 0 | 0.1 ± 0.1 | 23 ± 14 |

Table 2 – Comparison of PT-based and SG-based physiological signal extraction parameters in volunteers

The number of SI readouts is repeated by the number of acquired echoes ($N_{TE}$), which was 8. The cardiac and respiratory signals reconstructed from the 8 different SG and one PT sources are compared by quantification of the following metrics: percentage of spectral power filtered out due to trajectory dependencies (see **Supporting Information Figure S4**), Pearson correlation coefficient w.r.t. PT, percentage of difference in binning w.r.t. PT, percentage of missed trigger points w.r.t. the reference ECG, and ECG trigger jitter. All reported metrics are given as a mean and standard deviation.

No differences in the chest respiratory positions could be observed, indicating similarity in respiratory signals used for binning. Despite the higher variability in cardiac motion characteristics, the effect on reconstructed images was minor, but visible in the water-only images identified at end-systole and at mid-diastole for SG $TE_1$, SG $TE_7$ and PT (**Figure 4c**). All three sets of images displayed overall good visual agreement and homogenous fat suppression in both the chest and the heart. However, closer inspection of the region containing the right coronary artery (RCA) showed that the images binned based on SG $TE_7$ signals were blurrier, with a slight loss of contrast in the RCA at diastole.

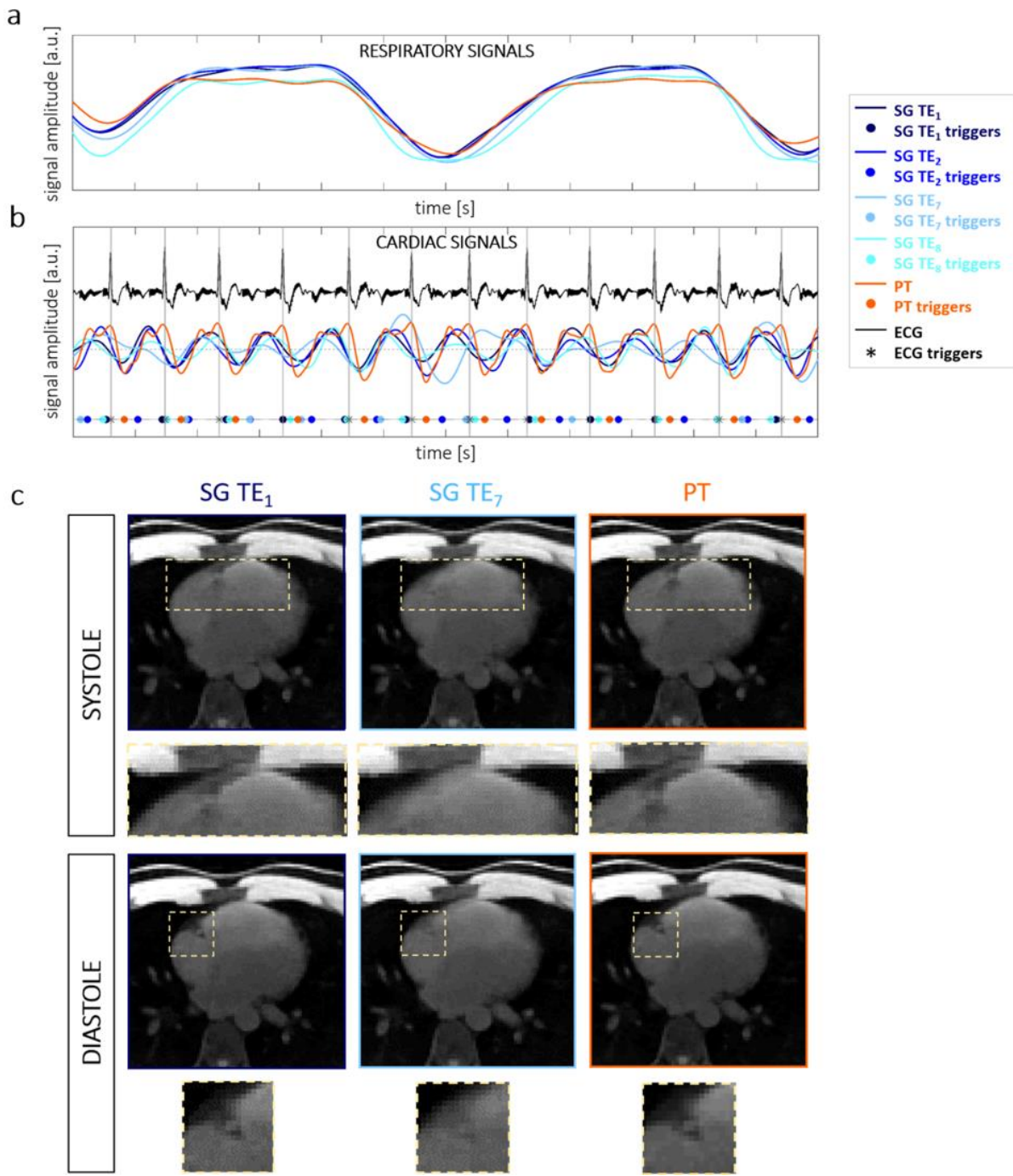

Figure 4 – Comparison of PT-based and SG-based physiological signal extraction in V1 and effect on water-only images

(a) Respiratory curves over a 12s time interval for healthy volunteer V1, obtained from PT data and from the SI projections of respective echo times $TE_1$, $TE_2$, $TE_7$ and $TE_8$.

(b) Corresponding cardiac signals over the same time interval. While the SG signals from the first two echoes show good correlation with PT, the signals from SG $TE_7$ and SG $TE_8$ present an offset w.r.t. PT. SG and PT trigger points extracted from each source are shown on the lower line, alongside with the ECG triggers for reference. ECG trigger points correspond to the R-wave, SG trigger points correspond to the zero-crossing (dotted line) of the extracted SG signal, and PT trigger points correspond to the local minima of the extracted PT signal. In V1, a main cardiac frequency $f_{CARD}$=1.03Hz was extracted from SG $TE_1$ and $TE_2$, while $f_{CARD}$=0.75Hz was reported with SG $TE_7$. The reference ECG yielded a mean cardiac cycle length of 1005ms (**Supporting Information Figure S5**), corresponding to a main cardiac frequency $f_{CARD}$=1.00Hz.

(c) Transverse mid-systolic and late-diastolic water-only images obtained from the proposed free-running sequence using the SG signals extracted from $TE_1$ (left), $TE_7$ (middle) and using PT signals (right). For both cardiac phases, images from all three physiological signal source are in good visual agreement. Close-ups of the regions containing the right coronary artery (RCA) show no difference between SG $TE_1$ and PT, but loss of contrast and additional blur at SG $TE_7$.

### 3.3 WATER-FAT SEPARATION AND PARAMETRIC MAPPING

The average post-processing time was 2h43min±7min for free-running and 14min37s±18s for triggered acquisitions.

Water-fat separation with graph cut was successful in all volunteers, without water-fat swaps or motion ghost artifacts. In the maps (**Figure 5**), no displacement of the expected static (chest, spine) regions was seen, nor ghosting, indicating that respiratory motion compensation was achieved and that cardiac motion compensation did not interfere with organs and tissue in the periphery of the FOV. The displacement of the fatty regions of the heart was clearly observed (**Figure 6**, animated GIF in **Supporting Information Video S1**).

Despite the lack of blood-to-myocardium contrast inherent to GRE imaging, the water-only images provided good visualization of cardiac anatomy, with complete absence of fatty tissue, enabling the visualization of the RCA(**Figure 5a**, top and **Figure 6c**). In the corresponding fat-only images (**Figure 5a**, bottom and **Figure 6d**), 3D visualization of pericardial fat was possible, particularly in coronal orientation around the right atrium and left ventricle. Sub-cutaneous fat-based tissues appeared brighter than cardiac fat tissues; such differences are more apparent in the co-registered quantitative FF and WF maps displayed with a color gradient (**Figure 5b**).

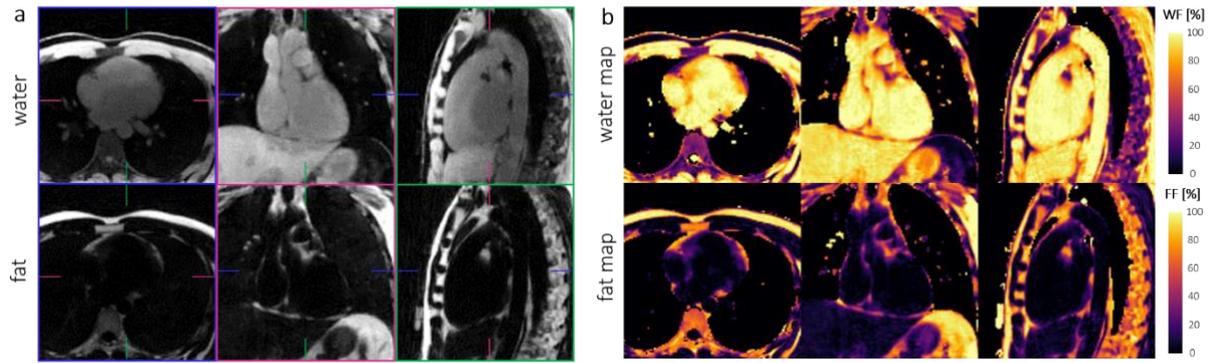

**Figure 5 – Water-fat separation and quantification obtained with the proposed framework**

On panel (**a**), water-only and fat-only images of healthy volunteer V1 are shown, as obtained after post-processing reconstructed 6D imaging volumes with a maximum-likelihood graph cut fitting algorithm. On panel (**b**), the corresponding parametric maps of water fraction and fat fraction are displayed. Slice position with respect to the three traditional MRI views (transversal, coronal and sagittal) is indicated on panel (**a**) by the colored lines.

In addition, the proposed framework produced cardiac and respiratory motion-resolved 3D maps of R2* (**Figure 6c, Supporting Information Video S1c**) and $B_0$ (**Figure 6f, Supporting Information Video S1f**). R2* values measured in the myocardium were within the expected range for healthy subjects (R2*<50Hz i.e. T2*>20ms). Elevated values (R2*>70Hz) were detected at the interfaces with the liver and the lungs. Compared to the FF and WF maps, the R2* maps displayed a higher standard deviation across the heart, with a more granular aspect, making cardiac motion less visible. The $B_0$ maps showed good homogeneity in the heart, without variations across the cardiac cycle. Off-resonance deviations were observed outside the heart, with deviations up to 100Hz seen in the liver. Changes in $B_0$ across the cardiac bins were only observed within air-filled regions (lungs and FOV periphery).

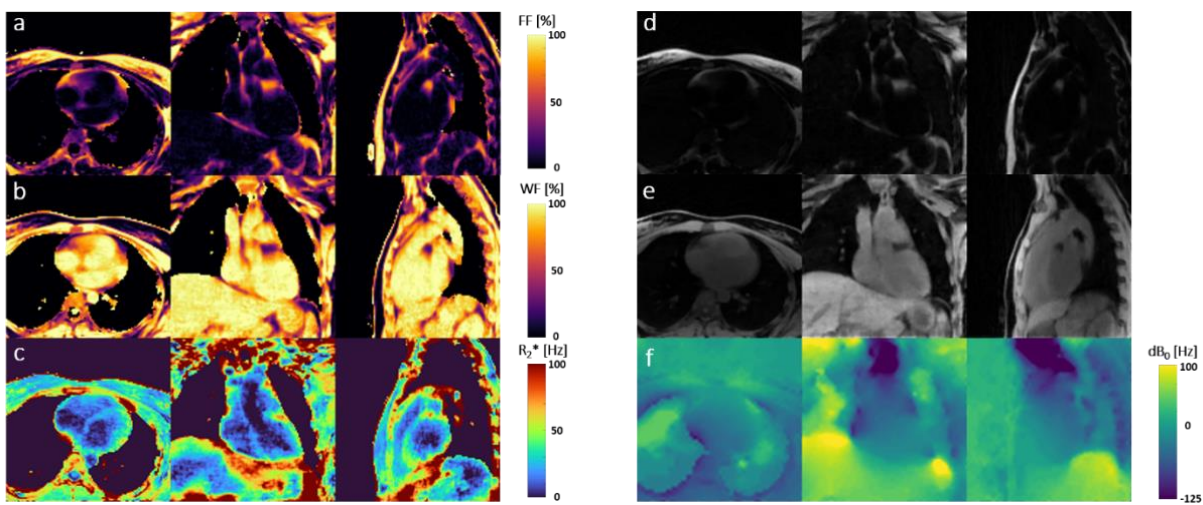

Figure 6 – **Respiratory and cardiac motion-resolved parametric maps in healthy volunteer V3: (a) fat fraction, (b) water fraction, (c) R2\* and (f) B$_0$, and separated fat-only (d) and water-only (e) images** *(see Supporting Information S1 for the animation)*

Transversal, coronal and sagittal views of the heart at end-expiration are displayed in panels (a), (b), (e), (d) and (f), while panel (c) displays a different set of slices, selected to highlight the myocardium delineation in the R2* maps. The animation presented in **Supporting Information Video S1** loops through the cardiac cycle.

Measurements in sub-cutaneous fat were not affected by motion binning, with constant FF measured throughout the cardiac cycle in all volunteers (**Figure 7a**). However, the FF measured in the pericardial fat varied across bins. A consistent pattern was found in all subjects, although to a different extent, with reduced FF at end-systole compared to mid-diastole (45.7±8.3% against 59.1±14.9% in V4, 59.0±16.5% against 66.0±10.0% in V10 after T1 bias correction). The average decrease in FF observed during end-systole across volunteers was 11.4±3.1% after T1 bias correction. Inspection of the corresponding B$_0$ maps (**Figure 7c**) revealed no deviations from one cardiac bin to the next, thus excluding B$_0$ inhomogeneities as cause of the observed FF variations. Higher standard deviations were reported in pericardial fat than in sub-cutaneous fat. The average FF measured in the ECG-triggered images was elevated w.r.t. the free-running diastolic images in 6 volunteers (**Figure 7a**). A Bland-Altman analysis showed a bias of -1.06% between the average pericardial FF measured in the free running maps identified as diastolic and the ECG-triggered maps, with 95% limits of agreement at [-8.70;6.58]% (**Supporting Information Figure S6**). Smaller standard deviations were observed in the ECG-triggered FF maps. Visually, the ECG triggered FF map had an apparent sharper delineation of the pericardial fatty tissue, where the free-running maps showed fatty regions spread over larger areas (**Figure 7b**).

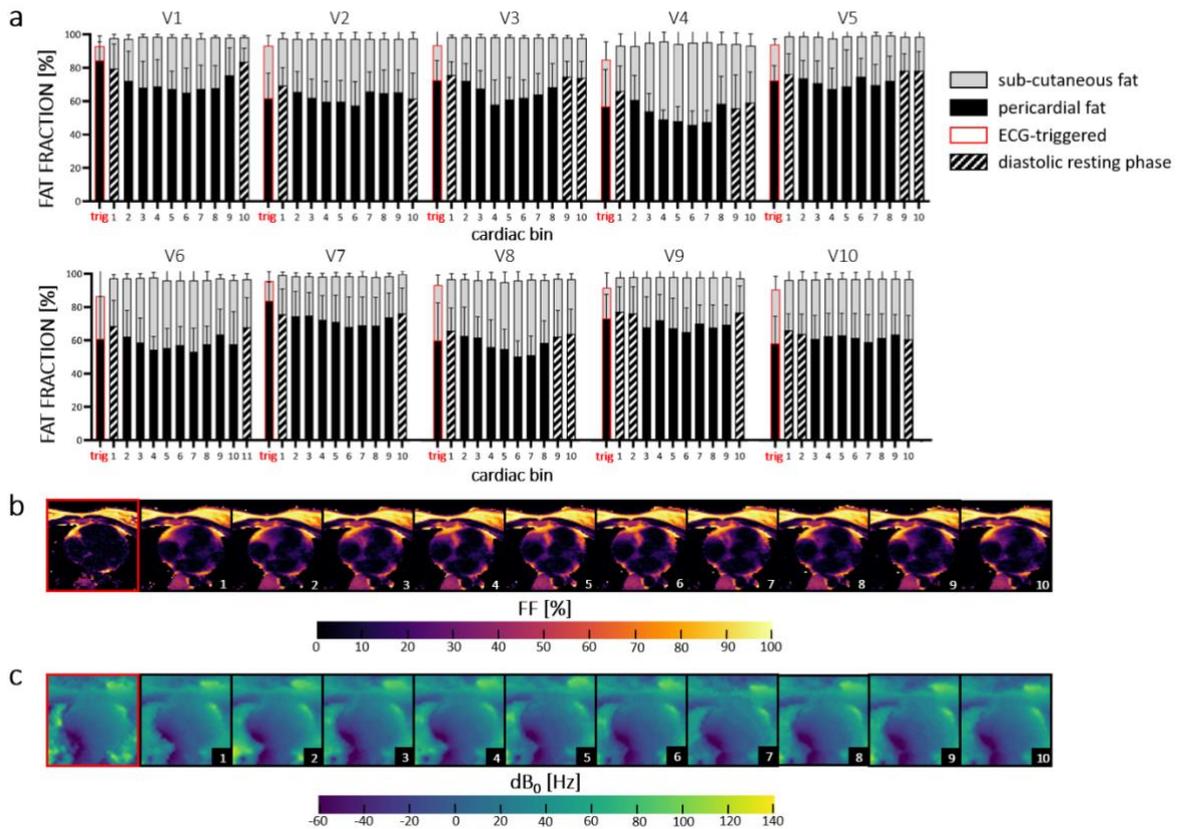

**Figure 7 – Fat quantification throughout the cardiac cycle, after T1 bias correction**

The proposed ME-GRE free-running framework allows quantitative visualization and measurement of cardiac fat content at each phase of the cardiac cycle.

(**a**) Average fat fraction and standard deviation measured in ROIs placed within two tissue types: sub-cutaneous fat and pericardial fat, at expiration. TR-adjusted T1 bias correction based on T1 estimates was performed according to the level of bias in the free-running and triggered sequence. In the free-running images, fat fraction was measured in each of the 10 (resp. 11 for volunteer V6) cardiac phases that were re-ordered to place end-diastole at the start (cardiac bin 1). Diastolic resting phase in the free-running images is identified with the white horizontal striped pattern. The matching ROIs reference measurements in ECG-triggered datasets (corresponding to diastolic resting phase) are shown in red borders on the left-hand side of each sub bar plot, with the label "trig".

(**b**) Transversal cardiac FF maps of healthy volunteer V3, at each cardiac bin as labelled in panel (**a**). The displacement of pericardial fatty regions can be followed throughout the cycle. The corresponding slice of the FF map obtained with the 5-echo ECG-triggered protocol is shown on the left, with red borders.

(**c**) Corresponding $B_0$ maps, which show deviations from the main magnetic field due to system-level imperfections and susceptibility effects. $B_0$ mapping constitutes an important step of fat quantification techniques with ME-GRE, as the off-resonance caused by the presence of fat has to be decoupled from other sources of off-resonance to be quantified accurately. The corresponding slice of the $B_0$ map obtained with the 5-echo ECG-triggered protocol is shown on the left, with red borders.

Although the free-running FF maps obtained from processing $N_{TE}$=4 or $N_{TE}$=8 visually agreed (**Figure 8a**), significant differences were found in both sub-cutaneous ($P<0.0001$) and pericardial fat ($P<0.01$) FF measurements when $N_{TE}$ is reduced (**Figure 8c**). The influence of a reduced $N_{TE}$ on R2* quantification was directly discernible from the maps (**Figure 8b**). Under-sampling in the echo dimension resulted in noisy maps, disrupting the visualization of anatomy such as myocardium delineation that was present with $N_{TE}$=8.

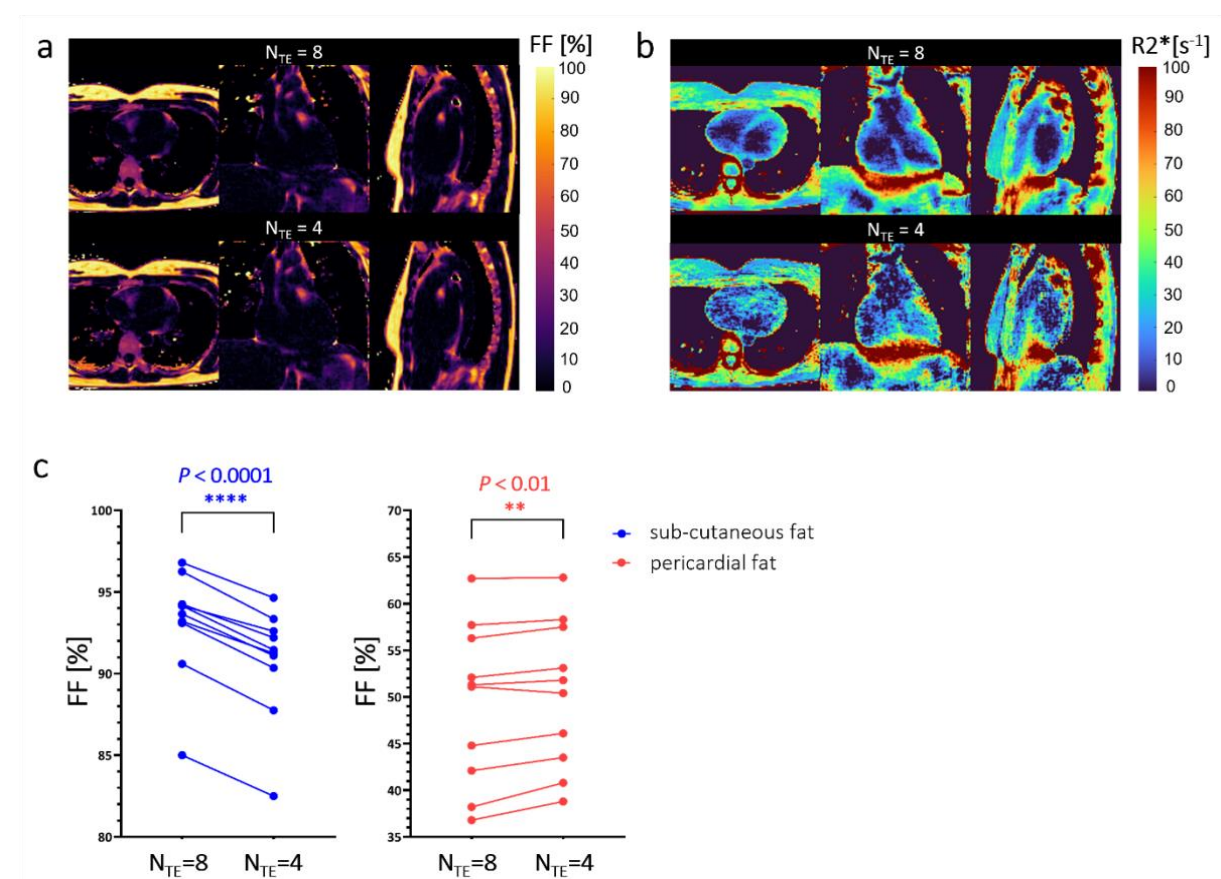

**Figure 8 – Influence of $N_{TE}$ on fat quantification**

(**a**) 3D fat fraction maps obtained with the proposed ME-GRE free-running framework, using $N_{TE}$=8 (top) and $N_{TE}$=4 (bottom) echoes for chemical species separation and quantification, in volunteer V2. Perfect visual agreement between both sets of images can be observed.

(**b**) 3D R2* maps obtained with the proposed ME-GRE free-running framework, using $N_{TE}$=8 (top) and $N_{TE}$=4 (bottom) echoes in volunteer V2. If a basic delineation of the myocardium muscle is visible in all three orientations with 8 echoes, the undersampled R2* maps lose this sighting due to high granularity.

(**c**) Average fat fraction in sub-cutaneous and pericardial fat ROIs measured in 10 selected slices, for all volunteers. The same ROIs were used in both fully sampled and undersampled maps. A paired parametric t-test reveals a highly significant difference ($P<0.0001$) in sub-cutaneous fat measurements, with a consistent underestimation of the fat content measured with $N_{TE}$=4 between volunteers. In pericardial fat, the influence of $N_{TE}$ on fat quantification was also detected, but at a lower level of significance ($P<0.01$).

## 4. DISCUSSION

This study demonstrates the feasibility of free-running whole-heart ME-GRE based fat fraction mapping using retrospective motion resolution based on PT signals and XD-GRASP. The free-running FF mapping approach with PT permits the collection of an unlimited number of echoes, thus benefiting mapping accuracy. In the current proof-of-concept study, $N_{TE}=8$ echoes were acquired in a relatively short scan time (6:15min) for 3D whole-heart coverage, while only $N_{TE}=5$ echoes could be acquired in a corresponding ECG-triggered acquisition with similar TA. With the proposed free-running ME-GRE framework, signal decay as a result of T2* relaxation provides the only limitation on $N_{TE}$.

Cardiac motion signals extracted from PT showed an improved correlation with the reference ECG trace than SG signals, confirming previous findings[50]. PT navigation strategies have several advantages compared to SG strategies. First, PT navigation provides a constant and higher sampling frequency (2kHz). Second, PT signals are insensitive to the underlying trajectory, unlike SG signals which require a correction for trajectory-dependent signal modulation[47]. Finally, scan time could be further shortened by removing the SI projections. In the current sequence design, the frequency of SI readouts decreases with increasing $N_{TE}$, thus affecting temporal resolution: with $N_{TE}=8$, the SG sampling frequency was 2.67Hz, which may hinder cardiac rhythm detection. Although it was not observed to be a limitation in the current healthy volunteer study, it is a potential hurdle for signal extraction in patient populations with abnormal or altered cardiac rhythm. In such cases, PT would remove the need for signal extrapolation, which is otherwise required. Furthermore, the analysis of SG signals extracted from various echoes demonstrated that some frequency components, in particular towards the end of the echo train where SNR is decreased, do not accurately depict cardiac rhythm and therefore introduce motion blurring. However, despite differences in cardiac signals from the PT and SG sources, the effect on image quality after water-fat separation was minor at (2mm)$^3$ spatial resolution. Based on these findings, PT navigation may especially benefit methods where ≥12 echoes are required, such as T2* or complex fat models. Alternative PT-like methods, such as Beat-PT[61], have shown promising results for accurate triggering and could be considered to further improve the cardiac signal extraction. Otherwise, self-gating extraction techniques based on k-space center information could be considered as a way to increase the frequency of motion sampling[38,41,42,62].

The use of XD-GRASP in this study allowed to recover anatomy that is otherwise barely distinguishable with a simple regridded reconstruction of the heavily undersampled ME-GRE radial data. The reconstruction framework also maintained the contrast changes expected from one echo to another in ME-GRE, and images were in agreement with the reference ECG-triggered ones. While other free-running studies make use of bSSFP readouts and report strong aliasing due to bright adipose tissue in non-fat suppressed highly accelerated scans[38], our reconstructed ME-GRE images did not exhibit such artifacts. The current sequence design did not allow for undersampling in the echo dimension, as was done previously[63], where each subsequent echo had a different trajectory. This approach, as well as rosette-like trajectories[64] where the acquisition of multiple "petals" mimics a multiecho scan, could be exploited to reduce scan time.

The free-running ME-GRE framework enabled quantitative fat measurements across the cardiac cycle, which allows to select a preferred frame for various adipose tissue characterization[65], and in the current volunteer study, a variable FF pattern was consistently observed as a function of time, indicating an heterogeneity of the pericardial fatty tissue. This result suggests that motion-averaged or frozen-motion (i.e. triggered) visualizations might only allow for partial or incomplete tissue characterization, a finding that can be linked to similar reports on the effect of respiratory motion on R2* quantification in the liver[66,67]. It remains to be determined whether the observed variation is physiological, or whether ROI drawing on (2mm)$^3$ spatial resolution maps might have affected the results. Because no $B_0$ variations were detected in the heart across the cardiac phases, the observation is not related to $B_0$ main magnetic field. The $B_0$ maps produced by the framework mainly serve to separate water and fat, but could have the potential to detect and correct errors or artifacts in images acquired during the same CMR exam that are sensitive to off-resonance[68]. Despite the visual agreement between FF maps obtained with 4 and 8 echoes, significant differences in FF in both static and moving tissues suggest that higher echo sampling benefits mapping accuracy. This is further corroborated by the difference in sub-cutaneous fat FF measurements between the 5-echo ECG-triggered acquisition and the 8-echo free-running acquisition, even after TR-adjusted correction from T1 bias. If a reduced flip angle approacg is commonly used for T1 bias reduction, phantom results suggest that an a posteriori correction based on T1 estimates, while requiring extra information, might be better suited for the proposed framework, as it would allow maintaining higher SNR, following the conclusions of Yang *et al.*[20]. T1 bias could be circumvented with the use of emerging FF quantification methods which employ different ways to encode off-resonance[69], but are not applicable to ME-GRE data.

While the same reference fat spectral model was used throughout all the experiments in this study[56], the sampling of $N_{TE}$=8 echoes would allow to test different models, including ones with various amounts of peaks (theoretically up to $N_{TE}$-2=6 peaks[18,23]) or ones based on cardiac fatty tissue. Future investigations could therefore include the use of self-calibrated spectra, as reported by several studies[25,70]. Although not explored in the current study, the R2* quantification and the resulting R2* map suggest that a delineation of the myocardial muscle is possible. The current framework could therefore be used for the simultaneous assessment of myocardial iron overload[71,72], by extending it to bi-exponential relaxation models for a refined estimation[73,74].

Although the water-fat quantification methods (T2*-IDEAL, graph cuts algorithm) have been validated thoroughly in phantoms[24,75], a ground truth measurement for pericardial FF could not be performed in this volunteer study. Besides invasive biopsies and PDFF measured with MRS, which are focal measures, there currently exists no noninvasive reference measurement for whole-heart FF *in vivo*.

To maintain a reasonable scan time, the data was heavily undersampled and therefore the binning limited to 10-11 cardiac and 2 respiratory phases. Consequently, high regularization was used in the CS reconstruction, allowing to suppress residual sampling artifacts but at the expense of slightly compressing the motion, as exhibited by the Likert scores in motion resolution (**Supporting Information Table S2**). This limitation could potentially be overcome by performing motion compensation in the respiratory domain[76]. Subsequently, this would allow for a cardiac motion-resolved reconstruction with an increased number of cardiac bins[77] and the recovery of a higher Nyquist factor. Furthermore, the addition of motion fields within the reconstruction may provide improved image sharpness and visualization of cardiac fat[26]. The use of a motion-consistent approach based on clustering could also be integrated in this type of golden-angle acquisition[78].

An additional challenge is the absence of large volumes of fat tissue in our volunteers, which at $(2.0mm)^3$ resolution restricts the number of voxels available for quantitative and statistical analysis, and makes segmentation more difficult. Future work in patients (displaying wider ranges of FF) would allow for a comparison with invasive biopsies, and may help determine whether the proposed framework, at increased spatial resolution, allows for a distinction of epicardial vs. pericardial fat, within the frame of coronary artery disease assessment[79,80].

By incorporating an additional echo dimension within the free-running framework, the present study constitutes a preliminary step towards easier access to whole-heart fat quantification, while the presence of cardiac fat as an imaging biomarker undergoes early stage validation.

## 5. CONCLUSION

In this study, an MRI framework incorporating a free-running ME-GRE sequence, integrated PT navigation, a robust CS reconstruction and a multi-peak fat fraction mapping routine was proposed for whole-heart water-fat separation and quantification. The framework combines ease-of-use, robustness to motion and whole-organ fat quantification without compromising on the number of collected echoes, in a scan time of 6:15min.

# DATA AVAILABILITY STATEMENT

All 10 volunteer datasets, consisting of 1) the free-running and ECG-triggered raw data files, 2) the raw Pilot Tone signals, 3) the extracted respiratory and cardiac signals based on processing of the raw Pilot Tone signals, and 4) the corresponding time stamps, are publicly available from the following public repositories:

Part 1 (V1-V5): https://zenodo.org/record/7621356#.Y-ZFaXbMLcu

Part 2 (V6-V10): https://zenodo.org/record/7615780#.Y-ZFfXbMLct

Volunteer data was collected and approved for open research sharing under the local ethics authorization CER-VD 2021-00708 (Lausanne, Switzerland).

MatLab scripts to read the raw data, compute the 3D trajectory and read the raw and extracted Pilot Tone signals are available from the following public repository:

https://github.com/QIS-MRI/ReadDataAndTrajectory_FreeRunningFatFractionHeart

The compressed-sensing based motion-resolved image reconstruction contains proprietary information that cannot be made available.

# ACKNOWLEDGEMENTS

We acknowledge the use of the Fat-Water Toolbox (http://ismrm.org/workshops/FatWater12/data.htm) for some of the results shown in this article. The authors thank Diego Hernando for the valuable discussions and input on this work. This study was supported by funding received from the Swiss National Science Foundation (grants #PCEFP2_194296 and #PZ00P3_167871 to JB, grant #PZ00P3_202140 to CWR, grants #320030B_201292 and #320030_173129 to MS), the Unil Bourse Pro-Femmes (JB), the Emma Muschamp Foundation (JB) and the Swiss Heart Foundation (grant #FF18054 to JB).

Whole-heart motion-resolved free-running fat fraction mapping – Adèle L.C. Mackowiak *et al.* – 202263. Benkert T, Feng L, Sodickson DK, Chandarana H, Block KT. Free-breathing volumetric fat/water separation by combining radial sampling, compressed sensing, and parallel imaging. Magn. Reson. Med. 2017;78:565–576 doi: 10.1002/mrm.26392.

64. Liu Y, Hamilton J, Eck B, Griswold M, Seiberlich N. Myocardial T1 and T2 quantification and water–fat separation using cardiac MR fingerprinting with rosette trajectories at 3T and 1.5T. Magn. Reson. Med. 2021;85:103–119 doi: 10.1002/mrm.28404.

65. Daudé P, Ancel P, Confort Gouny S, et al. Deep-Learning Segmentation of Epicardial Adipose Tissue Using Four-Chamber Cardiac Magnetic Resonance Imaging. Diagnostics 2022;12:126 doi: 10.3390/diagnostics12010126.

66. Zhong X, Armstrong T, Nickel MD, et al. Effect of respiratory motion on free-breathing 3D stack-of-radial liver relaxometry and improved quantification accuracy using self-gating. Magn. Reson. Med. 2020;83:1964–1978 doi: 10.1002/mrm.28052.

67. Schneider M, Benkert T, Solomon E, et al. Free-breathing fat and $R_2^*$ quantification in the liver using a stack-of-stars multi-echo acquisition with respiratory-resolved model-based reconstruction. Magn. Reson. Med. 2020;84:2592–2605 doi: 10.1002/mrm.28280.

68. Kellman P, Herzka DA, Arai AE, Hansen MS. Influence of Off-resonance in myocardial T1-mapping using SSFP based MOLLI method. J. Cardiovasc. Magn. Reson. 2013;15:63 doi: 10.1186/1532-429X-15-63.

69. Rossi GM, Hilbert T, Mackowiak AL, Pierzchała K, Kober T, Bastiaansen JA. Fat fraction mapping using bSSFP Signal Profile Asymmetries for Robust multi-Compartment Quantification (SPARCQ). 2020 doi: 10.48550/arXiv.2005.09734.

70. Reeder SB, Robson PM, Yu H, et al. Quantification of hepatic steatosis with MRI: The effects of accurate fat spectral modeling. J. Magn. Reson. Imaging 2009;29:1332–1339 doi: 10.1002/jmri.21751.

71. Triadyaksa P, Oudkerk M, Sijens PE. Cardiac $T_2^*$ mapping: Techniques and clinical applications. J. Magn. Reson. Imaging 2020;52:1340–1351 doi: 10.1002/jmri.27023.

72. Mavrogeni S. Evaluation of myocardial iron overload using magnetic resonance imaging. Blood Transfus. 2009;7:183–187 doi: 10.2450/2008.0063-08.

73. Juras V, Apprich S, Zbýň Š, et al. Quantitative MRI analysis of menisci using biexponential $T_2^*$ fitting with a variable echo time sequence. Magn. Reson. Med. 2014;71:1015–1023 doi: 10.1002/mrm.24760.

74. Positano V, Salani B, Pepe A, et al. Improved $T_2^*$ assessment in liver iron overload by magnetic resonance imaging. Magn. Reson. Imaging 2009;27:188–197 doi: 10.1016/j.mri.2008.06.004.

75. Hernando D, Sharma SD, Aliyari Ghasabeh M, et al. Multisite, multivendor validation of the accuracy and reproducibility of proton-density fat-fraction quantification at 1.5T and 3T
29

# SUPPORTING INFORMATION

**Supporting Information Table S1: Heart rate detection limitations**

| $N_{TE}$ | 1 | 2 | 3 | 4 | 5 | 6 | 7 | 8 |
|---|---|---|---|---|---|---|---|---|
| $f_{SI}$ [Hz] | 17.02 | 9.63 | 6.71 | 5.15 | 4.18 | 3.51 | 3.04 | 2.67 |
| $HR_{lim}$ [bpm] | 501 | 288 | 201 | 154 | 125 | 105 | 91 | 80 |

Increasing the number of echoes prolongs the duration of one radial phyllotaxis interleave, and thus prolongs the spacing between two superior-inferior (SI) segments with the same echo time TE. The higher the number of collected echoes $N_{TE}$, the longer this spacing and therefore the smaller the frequency of SI sampling $f_{SI}$. Using the self-gating approach, $f_{SI}$ determines the maximum heart rate $HR_{lim}$ that can be detected without aliasing due to undersampling (as per Nyquist-Shannon sampling theorem, $HR_{lim}$ [bpm] = $0.5*f_{SI}*60$). The values provided in this table were calculated using the sequence parameters of the proposed free-running acquisition ($TE_1$=1.25, ΔTE=2.05, nseg=22 radial segments).

**Supporting Information Figure S1: Comparison to Cartesian sampling trajectory in a phantom with controlled fat fraction vials**

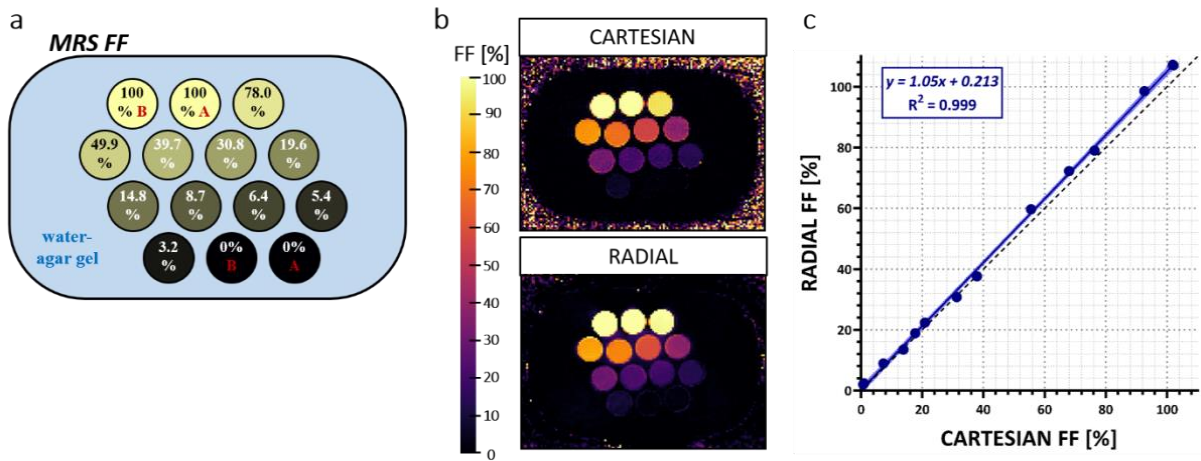

**a**. Custom fat-water phantom containing 14 Falcon vials of 50mL capacity placed in a plastic container. The bulk of the container was filled with a solidified agar-water solution of agar concentration 4%. The content of the vials consisted in different mixes of a water solution and a peanut-oil based solution, following a recipe from Bush *et al.*[81]. Volumes of peanut oil solution poured in the different vials were [0, 0, 1.25, 2.5, 3.75, 5, 7.5, 10, 15, 20, 25, 37.5, 50, 50]mL, and then completed to 50mL with the water solution when necessary, to achieve intended fat fractions of [0, 0, 2.5, 5, 7.5, 10, 15, 20, 30, 40, 50, 75, 100, 100]% respectively. The final percentage of fat present in each vial, indicated on the diagram, was measured with $^1$H NMR spectroscopy in a pre-clinical 9.4T scanner (Varian Medical Systems, Inc.) and analysed using jMRUI4.0 software.

**b**. Fat fraction maps of the phantom obtained using the proposed 3D radial free-running ME-GRE sequence (bottom), and the same acquisition but using a Cartesian sampling trajectory (top). Sequence parameters were identical for both sequences (FOV=290mm³, spatial resolution of 2mm³, RF excitation α=12°, pixel bandwidth of 890Hz/px, $N_{TE}$=8, $\Delta TE$=2.05ms and $TE_1$=1.25ms) at the exception of a slightly higher TR for the Cartesian sequence ($TR_{radial}$=17.02ms and $TR_{Cartesian}$=17.20ms). The same mapping parameters for the graphcut fitting routine as used in the healthy volunteers experiments (reported in Section 2.1.5.) were used to obtain the phantom FF maps.

**c**. Linear regression plot of the average FF estimated in radial phantom map as a function of the Cartesian estimations. The average fat fraction was measured in the same 14 cylindrical ROIs drawn in the vials in both radial and Cartesian maps.

Whole-heart motion-resolved free-running fat fraction mapping – Adèle L.C. Mackowiak *et al.* – 2022

**Supporting Information Figure S2: 3D radial phyllotaxis trajectories**

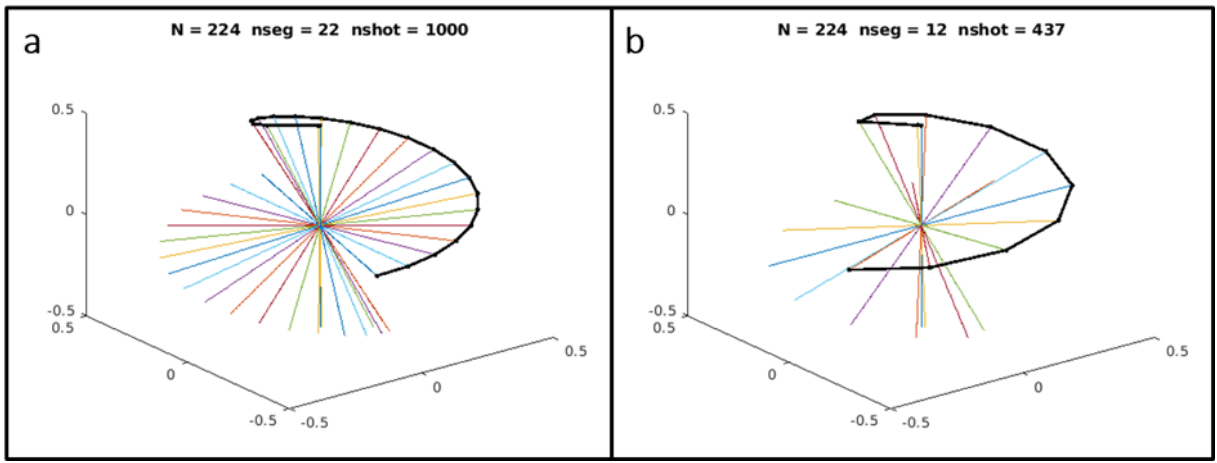

Panel (**a**) and (**b**) depict the phyllotaxis trajectory (see Piccini *et al.* [51]) and choices of "segments" and "shots" (or "interleaves") for the free-running and ECG-triggered acquisitions, respectively. In order to acquire enough radial lines to satisfy a 5% Nyquist sampling criterion, this pattern is repeated nshot times with a rotation of the golden angle between each shot.

**Supporting Information Figure S3: Motion-resolved image reconstruction with XD-GRASP**

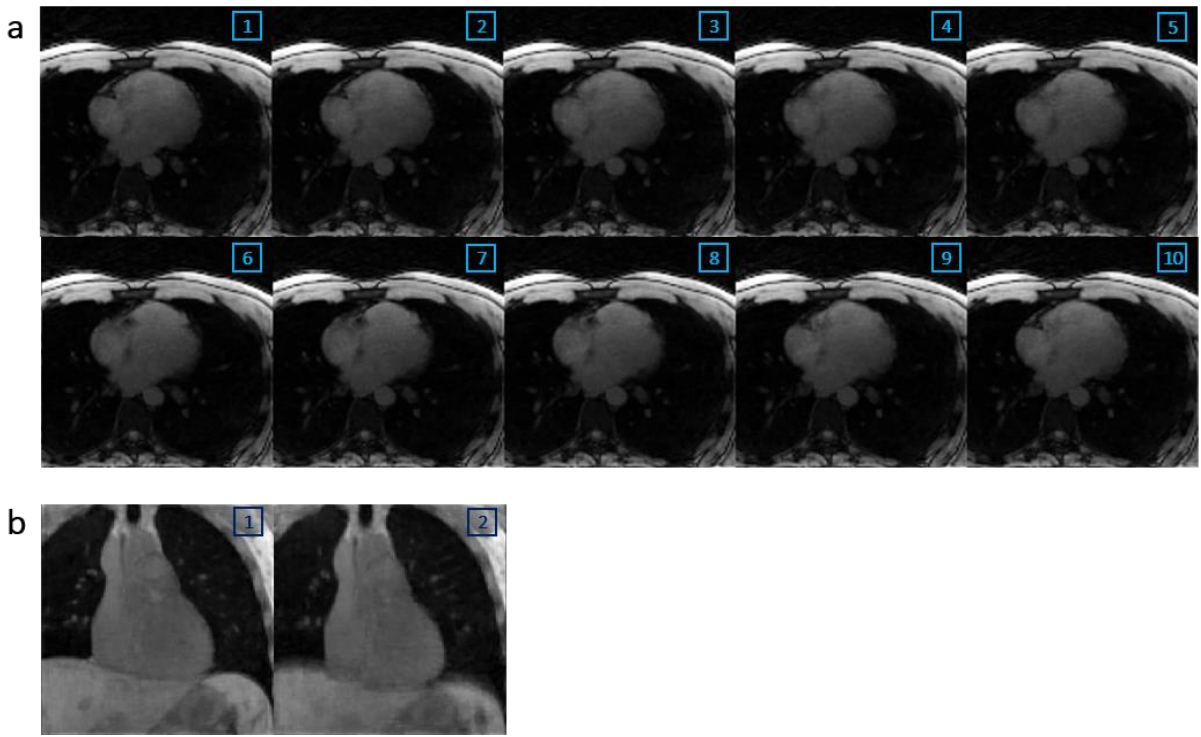

Different cardiac (**a**) and respiratory (**b**) phases of $TE_1$ reconstructed using XD-GRASP and the ADMM method. Data was sorted into different motion states using the Pilot Tone signal. Cardiac motion was binned into 10 states of duration 90ms, and respiratory motion was divided into 2 (expiration and inspiration) bins. Total variation regularization was applied along the cardiac temporal dimension with a weight of 0.0350, and along the respiratory temporal dimension with a weight of 0.0050.

**Supporting Information Figure S4: Filtering trajectory-dependent frequency components from self-gating signals**

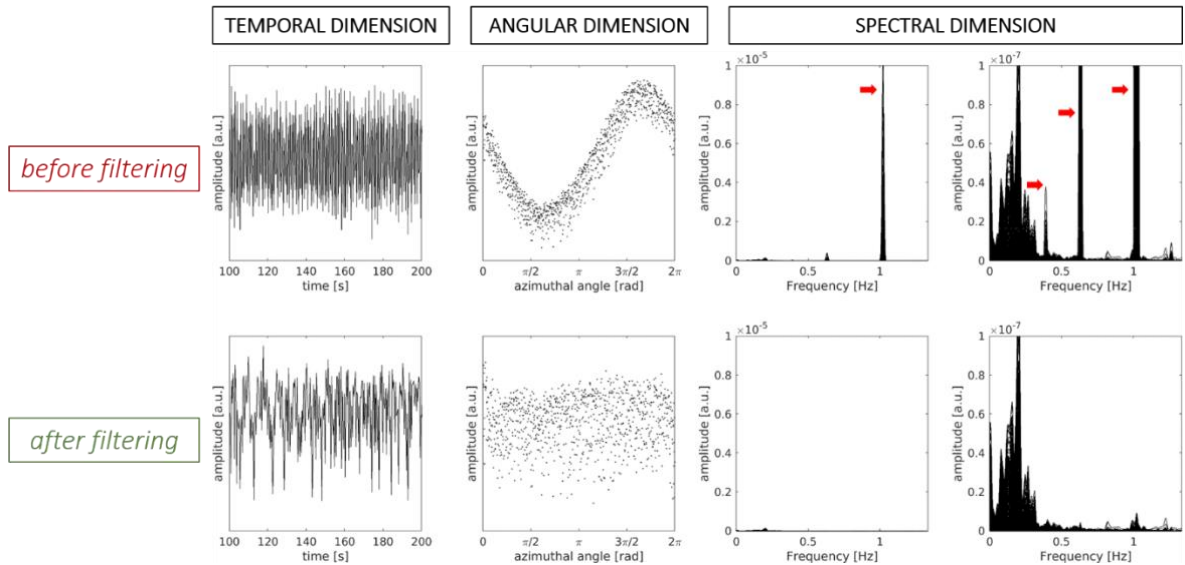

The signal amplitudes of the successively acquired superior-inferior (SI) readouts of the 3D phyllotaxis trajectory are modulated by physiological motion as well as trajectory imperfections caused by eddy currents and gradient delays. The trajectory-dependent modulation is detectable in the angular dimension (second column), where SI amplitudes are sorted and plotted as a function of the angle of their corresponding interleaf (shot) within the spiral pattern (see Di Sopra *et al.*[47]). In the spectral dimension (third and fourth columns), most of the trajectory-related power spectral density is concentrated in a sharp peak at 1Hz, and the expected lower-frequency components for respiratory motion are undistinguishable. Note that the third and fourth columns of the figure display the same signals in the spectral dimension, but with different scales in the y axis, as to show the over-whelming amplitude of trajectory-dependent components over physiological ones. The undesirable trajectory-dependent components of the signal are indicated in the spectral dimension by red arrows (third and fourth columns). After the use of a Butterworth IIR high-pass filter with stop band at 0.001rad$^{-1}$ and pass band at 0.04rad$^{-1}$, the modulation in the angular dimension is eliminated, and the remaining periodic components in the signal correspond to physiological motion (<0.4Hz for respiration and 1 Hz for cardiac motion) with realistic relative amplitudes in the spectral dimension.

**Supporting Information Table S2: Blinded scoring according to a 5-point Likert scale, averaged over n=10 volunteers**

|  | MOTION RESOLUTION | RESIDUAL SAMPLING ARTIFACTS | TISSUE DELINEATION |
|---|---|---|---|
| Water-only images | 2.2 ± 0.6 | 3.9 ± 0.4 | 3.2 ± 0.7 |
| Fat-only images | 2.8 ± 0.9 | 3.6 ± 0.5 | 3.5 ± 0.5 |
| Fat fraction maps | 2.2 ± 0.6 | 3.8 ± 0.4 | 3.6 ± 0.5 |

Two of the co-authors on this work (JY & CWR) knowledgeable in radial compressed sensing were asked to grade the water-only images, the fat-only images and the FF maps produced by the proposed framework. Three criteria were evaluated according to a 5-point Likert scale, following comparable methodology such as presented by Goldfarb *et al.*[34] and Jaubert *et al.*[31] The scoring was performed in consensus. The three criteria listed below were judged on expiratory images and maps, according to the following scales:

1) Motion resolution: 1=suppressed, 2=compressed, 3=mildly compressed, 4=good, 5=fully resolved.
2) Residual sampling artifacts: 1=severe, 2=high, 3=mild, 4=low, 5=absent.
3) Tissue delineation: 1=very poor, 2=poor, 3=acceptable/average, 4=good, 5=very good.

All three sets of images or maps received similar grades for all criteria evaluated. In the evaluation of the presence of residual sampling artifacts, the consensus grades were 3.9 for the water-images, 3.6 for the fat images and 3.8 for the maps. Tissue delineation received an overall slightly above average score with 3.2 for the water images, 3.5 for the fat images and 3.6 for the maps. All three sets of images however scored worse in motion resolution, with an average of 2.2 for water images and fat fraction maps, and 2.8 for the fat images. The overall evaluation from the experts was that while visualization of fat structures was rendered possible, especially given the minimal presence of undersampling (streaking) artifacts, the regularization most likely had compressed the motion.

It is to be noted that, as mentioned in other fat fraction mapping studies[82], FF quantification is not clinically performed yet therefore there is currently a lack of expertise reading and judging of the quality of these images.

**Supporting Information Figure S5: Linear regression of Pilot Tone vs ECG mean estimated cardiac cycle length in 10 volunteers**

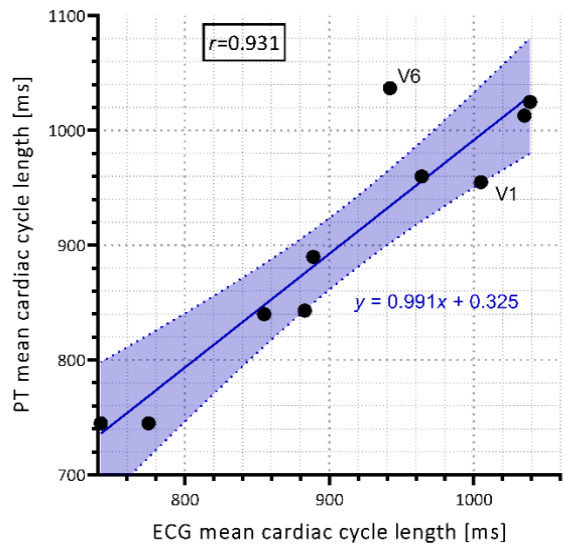

The 95% confidence interval on the linear regression, estimated at *y*=0.991*x*+0.325, is indicated by the shaded blue area. A Pearson's correlation analysis yielded a coefficient of *r*=0.931. Healthy volunteer V1 for which signals and images from the self-gating multiecho study were displayed (**Figure 4**) is indicated. Healthy volunteer V6, which showed the highest deviation with respect to the ECG, is also indicated. All statistical analyses were performed in GraphPad Prism v.8.0.2.

**Supporting Information Video S1: Proof-of-concept respiratory and cardiac motion-resolved parametric maps in healthy volunteer V3: (a) fat fraction, (b) water fraction, (c) R2\* and (d) $B_0$, and separated fat-only (e) and water-only (f) images**

The animation shows the transversal, coronal and sagittal views of the heart at the expiratory phase of the respiratory cycle, and loops through the cardiac cycle divided into 10 frames (i.e. cardiac bins in the reconstruction pipeline). The same slices are displayed in panels (**a**), (**b**), (**e**), (**d**) and (**f**), while panel (**c**) displays a different set of slices, selected to highlight the myocardium delineation in the R2\* maps.

**Supporting Information Figure S6: Bland-Altman (a) and regression (b) analysis of fat content measured in free-running images identified as diastolic resting phase, versus ECG-triggered data, after T1 bias correction**

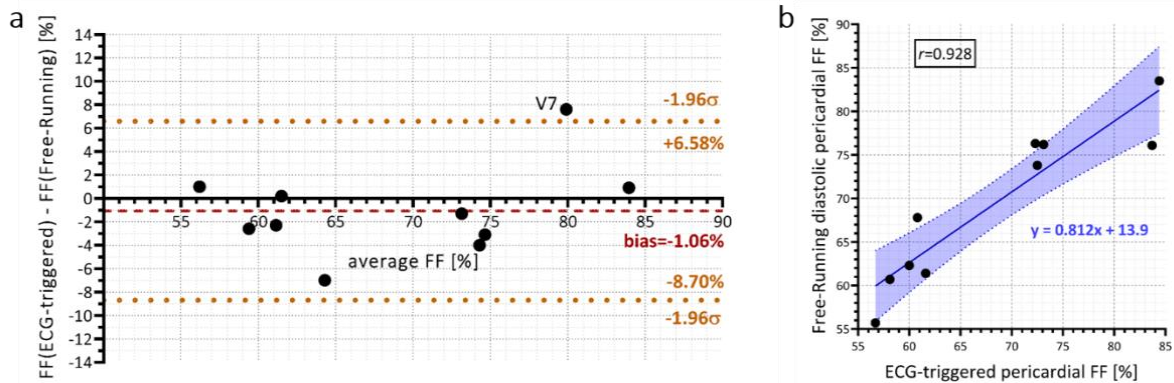

Pericardial fat fraction was measured in ECG-triggered maps and in free-running maps identified as diastolic. After a correction based on T1 estimates (see **Figure 2**), the Bland-Altman analysis yielded a bias of -1.06% (red dashed line). The limits of agreement at the 95% confidence level are indicated by the orange dotted lines. The average deviation observed was large in volunteer V7. The regression analysis yielded a curve $y=0.812x+13.9$. 95% confidence bounds are indicated by the shaded blue area. A Pearson's correlation analysis yielded a coefficient of $r=0.928$. All statistical analyses were performed in GraphPad Prism v.8.0.2.

**Supporting Information Table S3: Peak-to-peak respiratory amplitudes**

| physiological signal source | V1 | V2 | V3 | V4 | V5 | V6 | V7 | V8 | V9 | V10 | average over volunteers |
|---|---|---|---|---|---|---|---|---|---|---|---|
| SG TE$_1$ | 0.786 ± 0.074 | 0.284 ± 0.161 | 0.614 ± 0.127 | 0.476 ± 0.115 | 0.602 ± 0.103 | 0.672 ± 0.066 | 0.553 ± 0.144 | 0.714 ± 0.062 | 0.538 ± 0.109 | 0.823 ± 0.058 | 0.606 ± 0.149 |
| SG TE$_2$ | 0.811 ± 0.078 | 0.368 ± 0.124 | 0.656 ± 0.130 | 0.488 ± 0.119 | 0.647 ± 0.100 | 0.690 ± 0.075 | 0.461 ± 0.128 | 0.744 ± 0.075 | 0.624 ± 0.104 | 0.838 ± 0.041 | 0.633 ± 0.145 |
| SG TE$_3$ | 0.820 ± 0.064 | 0.535 ± 0.117 | 0.665 ± 0.132 | 0.536 ± 0.122 | 0.660 ± 0.103 | 0.680 ± 0.077 | 0.548 ± 0.137 | 0.748 ± 0.068 | 0.633 ± 0.104 | 0.848 ± 0.043 | 0.668 ± 0.106 |
| SG TE$_4$ | 0.837 ± 0.065 | 0.704 ± 0.133 | 0.756 ± 0.125 | 0.599 ± 0.112 | 0.719 ± 0.096 | 0.648 ± 0.070 | 0.480 ± 0.123 | 0.748 ± 0.062 | 0.670 ± 0.101 | 0.903 ± 0.031 | 0.706 ± 0.113 |
| SG TE$_5$ | 0.860 ± 0.049 | 0.761 ± 0.136 | 0.755 ± 0.128 | 0.633 ± 0.114 | 0.718 ± 0.096 | 0.655 ± 0.070 | 0.513 ± 0.141 | 0.767 ± 0.064 | 0.689 ± 0.101 | 0.879 ± 0.034 | 0.723 ± 0.103 |
| SG TE$_6$ | 0.849 ± 0.048 | 0.753 ± 0.121 | 0.789 ± 0.133 | 0.649 ± 0.118 | 0.681 ± 0.094 | 0.729 ± 0.070 | 0.483 ± 0.106 | 0.761 ± 0.057 | 0.736 ± 0.101 | 0.929 ± 0.028 | 0.736 ± 0.113 |
| SG TE$_7$ | 0.911 ± 0.039 | 0.793 ± 0.127 | 0.782 ± 0.131 | 0.663 ± 0.108 | 0.759 ± 0.092 | 0.736 ± 0.069 | 0.575 ± 0.131 | 0.775 ± 0.060 | 0.730 ± 0.096 | 0.892 ± 0.029 | 0.762 ± 0.093 |
| SG TE$_8$ | 0.886 ± 0.039 | 0.785 ± 0.087 | 0.784 ± 0.136 | 0.658 ± 0.111 | 0.792 ± 0.088 | 0.677 ± 0.077 | 0.665 ± 0.114 | 0.787 ± 0.060 | 0.774 ± 0.098 | 0.930 ± 0.026 | 0.774 ± 0.085 |
| Pilot Tone | 0.642 ± 0.098 | 0.166 ± 0.143 | 0.616 ± 0.125 | 0.499 ± 0.132 | 0.600 ± 0.138 | 0.764 ± 0.091 | 0.492 ± 0.141 | 0.816 ± 0.057 | 0.622 ± 0.121 | 0.823 ± 0.058 | 0.604 ± 0.183 |

The average and standard deviation of the peak-to-peak amplitude of the extracted respiratory signal measured in all 10 volunteers and all sources. The last column indicates the average value over volunteers. Note that these values are expressed in arbitrary unit (a.u.) as the signal extraction pipeline used in this work does not translate the signals into a physical displacement.

**Supporting Information Figure S7: Respiratory curves and binning in two volunteers**

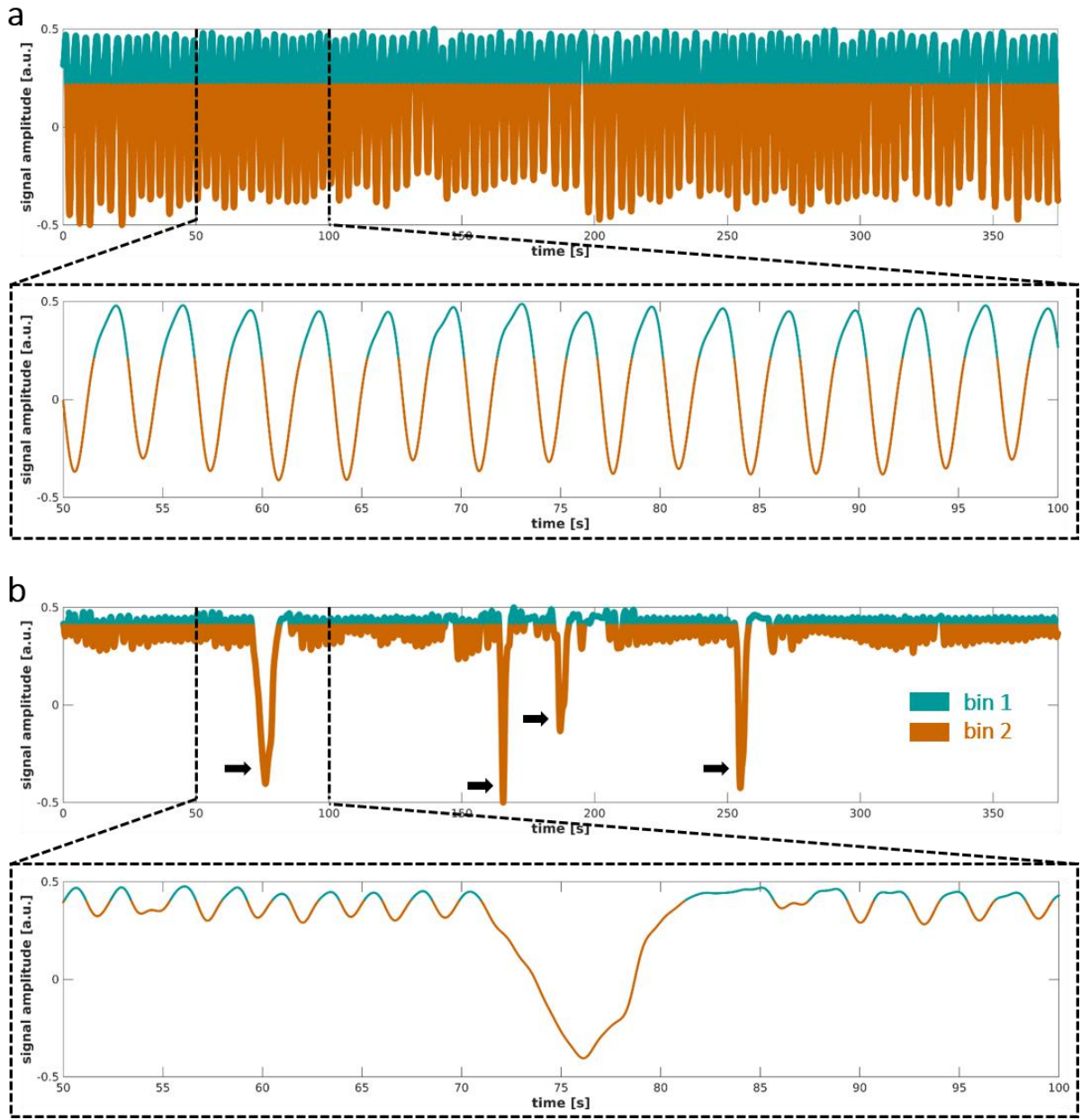

The respiratory signals from two healthy volunteers (V8 in panel **a** and V2 in panel **b**) are shown with the binning repartition. The physiological signal curves are expressed in arbitrary units. Images and maps showed in this paper (unless indicated otherwise) were reconstructed from data in bin 1, which correspond to the expiration state. Volunteer V8 had a very consistent respiratory pattern over the entire acquisition (**a**), while the respiratory signal curve from volunteer V2 showed large negative peaks (**b**, arrows), indicative of deep inspirations.